\documentclass[twocolumn,showpacs,10pt,aps,prd]{revtex4-1}
\usepackage[dvips]{graphicx}
\usepackage{amsmath,amssymb}
\newcommand{\sumint}{\int\hspace{-5.1mm}\sum}

\newcommand{\hlambda}{\hat{\lambda}}
\newcommand{\LQCD}{\Lambda_{\text{QCD}}}
\newcommand{\muq}{\mu_{\text{q}}}
\newcommand{\Lag}{\mathcal{L}}
\newcommand{\ua}{\mathrm{U(1)_A}}

\newcommand{\MeV}{\;\text{MeV}}
\newcommand{\fm}{\;\text{fm}}
\newcommand{\rmi}{\mathrm{i}}
\newcommand{\rmd}{\mathrm{d}}
\newcommand{\rme}{\mathrm{e}}
\newcommand{\Nc}{N_{\mathrm{c}}}
\newcommand{\Nf}{N_{\mathrm{f}}}
\newcommand{\Tc}{T_{\mathrm{c}}}
\newcommand{\bB}{\boldsymbol{B}}
\newcommand{\bE}{\boldsymbol{E}}
\newcommand{\bgamma}{\vec{\gamma}}

\newcommand{\bp}{\vec{p}}
\newcommand{\feyn}[1]{
  \setbox0=\hbox{\ensuremath{#1}}
  \hbox to\wd0{\hbox to0pt{\hbox to\wd0{\hss/\hss}\hss}\box0}}

\def\di{\displaystyle}

\def\bg{\begin{eqnarray}\begin{array}{rcl}\displaystyle}
\def\eg{\end{array} &\di    &\di   \end{eqnarray}}
\def\bm#1{\begin{eqnarray}\begin{array}{#1}\di}
\def\bmo#1{\begin{eqnarray*}\begin{array}{#1}\di}
\def\bml#1#2{\begin{eqnarray}\begin{array}{#1}\label{#2}\di}
\def\bgo{\begin{eqnarray*}\begin{array}{rcl}\displaystyle}
\def\ego{\end{array} &\di    &\di \nonumber  \end{eqnarray*}}

\def\btensor#1#2{\renew\left#1\begin{array}{#2}\di}
\def\brtensor#1#2#3{\ren#3\left#1\begin{array}{#2}}
\def\botensor#1#2{\renew\left#1\begin{array}{#2}}
\def\etensor#1{\end{array}\right#1}

\def\eq#1{(\ref{#1})}

\def\Tr{{\rm Tr}}

\def\s0#1#2{\mbox{\small{$ \frac{#1}{#2} $}}}
\def\0#1#2{\frac{#1}{#2}}

\begin{document}

\title{Magnetic catalysis in hot and dense quark matter
and quantum fluctuations}

\author{Kenji Fukushima}
\affiliation{Department of Physics, Keio University,
             Kanagawa 223-8522, Japan}
\author{Jan M.\ Pawlowski}
\affiliation{Institut f\"{u}r Theoretische Physik,
             Universit\"{a}t Heidelberg,
             Philosophenweg 16, 69120 Heidelberg, Germany}

\begin{abstract}
  We analyze chiral symmetry breaking in quark matter in an external
  magnetic field at zero and finite temperature and quark chemical
  potential.  We first give a brief overview of analytic results
  within the mean-field approximation.  There the critical temperature
  for chiral restoration is increased by the magnetic field effect.
  Then we investigate the effects of matter and quantum fluctuations
  on the Magnetic Catalysis.  More specifically, we compute the
  critical coupling as a function of the magnetic field and the
  temperature for zero and finite quark chemical potential in the
  presence of quantum fluctuations.  As soon as a non-zero temperature
  and/or density is turned on, long-range correlations are screened
  and the critical coupling is no longer vanishing.  We extend our
  dynamical results beyond the leading-order bubble resummation which
  results in a non-local four-Fermi coupling.  This includes in-medium
  meson effects on the more quantitative level.
\end{abstract}
\maketitle

\section{Introduction}

The state of matter in a strong magnetic field $\bB$ still poses
interesting theoretical as well as experimental challenges.  In the
Landau quantization approach to these situations the energy dispersion
relation of charged particles is quantized along the directions
perpendicular to $\bB$.  There are a countless number of interesting
phenomena related to the Landau quantization such as the Quantum Hall
Effect in condensed matter physics~\cite{Laughlin:1981}, the Chiral
Magnetic Effect in the relativistic heavy-ion
collision~\cite{Kharzeev:2007jp,Fukushima:2008xe} and related
phenomena~\cite{Metlitski:2005pr,*Kharzeev:2007tn,*Kharzeev:2010gr},
the primordial magnetic field in cosmological phase
transitions~\cite{Vachaspati:1991nm,*Cheng:1994yr,*Baym:1995fk}.  In
the present work we concentrate on the Magnetic Catalysis,
\cite{Gusynin:1994re,*Gusynin:1994va,*Gusynin:1994xp,*Gusynin:1995nb},
in the context of hot and dense quark matter.

In relativistic heavy-ion collisions the state of matter undergoes a
phase transition or cross-over from the quark-gluon plasma of strongly
interacting quark and gluon (quasi)-particles to the hadronic phase
with chiral symmetry breaking and confinement.  Recently,
$\bB$-induced phenomena within these phase transitions are attracting
more and more theoretical and experimental interest.  This is
motivated by the recognition of the presence of extraordinary strong
$\bB$ in the early stage of non-central
collisions~\cite{Kharzeev:2007jp}.  The specific importance of $\bB$
over the electric field $\bE$ is that $\bB$ is not screened by charged
particles unlike $\bE$.  The strength of $\bB$ can get as large as
$|e\bB|\sim\LQCD^2$, while its life time is also characterized by
$\sim\LQCD^{-1}$ and thus is very small~\cite{Skokov:2009qp}.
Nevertheless, it is conceivable to anticipate observable $\bB$ effects
in the distribution of produced
particles~\cite{Tuchin:2010gx,*Marasinghe:2011bt} since the production
process occurs within the time scale of the strong interaction or even
faster.

By now model analyses according to the hydrodynamic description of the
heavy-ion collision suggest that the thermalization is rapidly
achieved before $\tau\sim 0.6\;\text{fm/c}$~\cite{Heinz:2004pj}.
Because $\bB$ decays not exponentially but by the power as a function
of time, the magnetic field strength is still comparable to the QCD
scale when the system gets thermalized.  Consequently it is a very
interesting question how the QCD phase transition is modified by the
strong $\bB$ effects.

Within the framework of effective model studies it was found that the
chiral phase transition is significantly delayed (i.e.\ the critical
temperature is increased) by $\bB$, while the deconfinement transition
or the Polyakov loop behavior is hardly
affected~\cite{Fukushima:2010fe,Mizher:2010zb,Skokov:2011ib}.  This
latter result on deconfinement, however, turned out to be inconsistent
with the lattice-QCD simulation~\cite{D'Elia:2010nq,Bali:2011qj}.
Even though there are some attempts to account for the significant
modification in the Polyakov loop behavior by introducing extra
couplings~\cite{Gatto:2010pt,Kashiwa:2011js}, the microscopic details
about the coupling between the gluonic dynamics and the external
magnetic field are not fully understood yet.  The missing coupling is
possibly attributed to screening by polarization effects, which has
the same origin as the failure of chiral model approaches to
baryon-rich matter~\cite{Schaefer:2007pw,Fukushima:2010is}.  There it
has been shown within QCD computations that the polarization effects
have a considerable impact on the location of the
confinement-deconfinement phase transition; see
Refs.~\cite{Braun:2009gm,Pawlowski:2010ht}.  Indeed, in
Ref.~\cite{Galilo:2011nh} it has been shown that the interplay between
strong electromagnetic and chromo-electric and chromo-magnetic fields
gives rise to polarization effects that catalyze the deconfinement
phase transition.  This already hints at the similarity between
large-$\bB$ effect and high-$\muq$ effect which may give a clue for
attacking the problems at high density, as emphasized in
Ref.~\cite{Fukushima:2011jc}.

In contrast to deconfinement, the $\bB$ effects on the chiral phase
transition is naturally understood from the qualitative effects
implied by the so-called Magnetic
Catalysis~\cite{Gusynin:1994re,*Gusynin:1994va,*Gusynin:1994xp,%
  *Gusynin:1995nb} which was first discovered in fermionic theory with
four-Fermi interaction, and later the argument was extended to QED as
well~\cite{Gusynin:1995gt,*Gusynin:1998zq}.  When $\bB$ is
sufficiently strong, the magnetic field plays a role as the catalysis
that induces a finite chiral condensate
$\langle\bar{\psi}\psi\rangle$.  Since chiral symmetry breaking is
enhanced by $\bB$, it is natural to anticipate a higher critical
temperature for chiral restoration than in the case with vanishing
$\bB$.  This expectation was indeed confirmed in recent model
studies~\cite{Fraga:2008qn} and indeed was observed already before the
discovery of the Magnetic Catalysis~\cite{Suganuma:1990nn}.  In recent
lattice-QCD simulation~\cite{Bali:2011qj} it was found that $\Tc$
significantly decreases for larger $\bB$, which can be interpreted as
a result of the entanglement with the Polyakov loop that is affected
by the magnetic screening~\cite{Fukushima-Kashiwa}.  Such screening
has been discussed in Ref.~\cite{Galilo:2011nh} for the gluonic
potential.  In this sense, as long as only the chiral sector is
concerned, the enhancement of the chiral condensate by $\bB$ is not
contradictory to Ref.~\cite{Bali:2011qj}.  Also, regarding another
possibility of the inverse Magnetic Catalysis at finite density, see
Ref.~\cite{Preis:2010cq}.

The purpose of this paper is to study the $\bB$ effects on the chiral
phase transition including quantum fluctuations.  These effects of
quantum fluctuations are quite naturally included with
renormalization group (RG) flows which also shed light on the physics
mechanisms at work (see
Refs.~\cite{Hong:1996pv,Skokov:2011ib,Scherer:2012nn} for related
works).  We calculate the $\beta$-function for the four-Fermi coupling
constant $\lambda_k$ as a function of an IR cutoff scale $k$.  At
large cutoff scales $k$ implying large energies we are in the chirally
symmetric phase with vanishing chiral condensate
$\langle\bar{\psi}\psi\rangle=0$.  Chiral symmetry breaking is
signaled by a singularity in the four-Fermi coupling related to the
appearance of massless pion (and sigma) propagation in the
intermediate state of the four-point vertex at vanishing momenta.
Hence by decreasing the cutoff scale $k$ the RG-flow of $\lambda_k$
hits this singularity at some point (see Ref.~\cite{Braun:2011pp} for
a recent comprehensive review).  In this paper we adopt a chiral model
for simplicity.  In this way we can get a simple formula for the
critical surface in space spanned by the four-Fermi coupling
$\lambda_k$, the temperature $T$, and the quark chemical potential
$\muq$ with and without a constant magnetic field $\bB$.  The
extension or embedding of the model results obtained here within the
RG approach to QCD is straightforward; see
Refs.~\cite{Pawlowski:2010ht,Braun:2011pp}.

This paper is organized as follows.  We first present a brief overview
of the mean-field results for the critical coupling in
Sec.~\ref{sec:MF}.  This part is used for introducing our notation and
it also has the benefit of allowing an easy comparison of the
mean-field arguments and results with those of the RG analysis in
Sec.~\ref{sec:RG}.  Finally we extend the RG analysis by including
meson effects beyond the leading-order bubble resummation to the
four-Fermi coupling in Sec.~\ref{sec:meson}.  As we shall see in
Sec.~\ref{sec:RG} there are three different kinds of Feynman diagrams
that contribute to the $\beta$-function of $\lambda_k$.  Two of them
tend to break chiral symmetry and are genuinely fermionic in nature.
The last one forms a box-type diagram that favors chiral restoration
and this box diagram contains intermediate meson states with finite
momentum insertion.  This additional resummation establishes the
well-known effect that fermionic fluctuations tend to break chiral
symmetry whereas mesonic ones tend to restore it.  We find that the box
diagram has only a minor effect on the Magnetic Catalysis when $\bB$
is large.  Section~\ref{sec:conclusion} is devoted to the conclusion
and outlook.

\section{Overview of the Mean-Field Calculation}
\label{sec:MF}

Here we recall spontaneous chiral symmetry breaking on the mean-field
level with and without magnetic field.  We intend to capture the
generic feature of chiral symmetry breaking in an effective
description.  Integrating-out the gluons in QCD as well as the quark
fluctuations with momenta larger than a ultraviolet (UV) cutoff scale
$\Lambda$ would lead to a chiral effective theory formulated in terms
of quarks with all kinds of interaction vertices; see
e.g.\ Refs.~\cite{Berges:1998nf,Braun:2009gm,Kondo:2010ts,Pawlowski:2010ht}.
Among these vertices only the point-like four-point interaction is
retained for simplicity, which defines the Nambu--Jona-Lasinio (NJL)
model.  In what follows we consider hot and dense quark matter with
three colors, $\Nc=3$, and two massless flavors, $\Nf=2$.  Then the
Lagrangian density at the UV scale $\Lambda$ reads
\begin{equation}
 \Lag_\Lambda = \bar{\psi}\,\rmi\feyn{\partial}\psi + \frac{1}{2}\,
  \bar{\psi}_i^{a\alpha} \psi_j^{b\alpha}\; \Gamma_{\Lambda}{}_{ijlm}^{abcd}\;
  \bar{\psi}_l^{c\beta} \psi_m^{d\beta} \;,
\label{eq:L}
\end{equation}
where $\alpha$ and $\beta$ run in color space, $a$, $b$, $c$, $d$ in
flavor space, and $i$, $j$, $l$, $m$ refer to the Dirac indices.  We
take the sum over repeated indices as usual.  For the concrete form of
$\Gamma_{\Lambda}{}_{ijlm}^{abcd}$ we assume an equal mixing of the
$\ua$-symmetric and $\ua$-breaking terms so that the iso-scalar
pseudo-scalar particle ($\eta_0$) and the iso-vector scalar particles
($\vec{a_0}$) decouple from low-lying spectra.  The remaining
interaction vertices contain the iso-scalar scalar channel ($\sigma$)
and the iso-vector pseudo-scalar ($\vec{\pi}$) channels only,
\begin{equation}
 \Gamma_{\Lambda}{}_{ ijlm}^{abcd} = \lambda_\Lambda\bigl[
   \delta_{ij}\delta_{lm}\delta^{ab}\delta^{cd}
  + (\rmi\gamma_5)_{ij}(\rmi\gamma_5)_{lm}
  (\tau^n)^{ab}(\tau^n)^{cd} \bigr] \;,
\label{eq:Gamma}
\end{equation}
where $\tau^n$'s are the Pauli matrices in flavor space and $n$ runs
from $1$ to $\Nf^2-1$ corresponding to pion degrees of freedom. 

We close with the remark that the Lagrangian~\eqref{eq:Gamma} was
introduced as already comprising all quantum effects of quark
fluctuations with momenta larger than the cutoff scale as well as all
gluonic fluctuations.  This implies that \eqref{eq:Gamma} should have
included terms depending on $\Lambda$ as well as the physical mass
scales of QCD ($\LQCD, m_{\rm quark}$) such that if introducing the
missing quantum fluctuations of the quarks, the result is
$\Lambda$-independent.  This necessity is avoided by assuming
$\Lambda$ to be such that these terms are minimized.  This gives a
physical meaning to $\Lambda$ and it is rather $\LQCD$ up to a
constant than a UV cutoff scale.

\subsection{Gap equation and the critical coupling constant}

In the mean-field approximation the four-Fermi interaction is
decomposed into the mean-field $\langle\bar{\psi}\psi\rangle$ and the
fluctuations about it.  We pick up only the contribution from the
non-crossing color structure, $\langle\bar{\psi}^\alpha
\psi^\alpha\rangle\,\bar{\psi}^\beta\psi^\beta$, but simply neglect
the crossing one like $\langle\bar{\psi}^\alpha
\psi^\beta\rangle\,\bar{\psi}^\beta\psi^\alpha$.  Such a treatment is
justified in the large-$\Nc$ limit in which the mean-field
approximation also becomes exact.  The constituent quark mass,
$M=-\lambda_\Lambda\langle\bar{\psi}\psi\rangle$, appears as a result
of the spontaneous chiral symmetry breaking, and the mean-field
Lagrangian density reads
\begin{align}\label{eq:MFlag}
 \Lag_{\text{MF}} &= \bar{\psi}\,\rmi\feyn{\partial}\psi
  + \lambda_\Lambda\langle\bar{\psi}\psi\rangle\, \bar{\psi}\psi
  - \frac{\lambda_\Lambda}{2}\langle\bar{\psi}\psi\rangle^2 \notag\\
  &= \bar{\psi} (\,\rmi\feyn{\partial}-M ) \psi
  - \frac{M^2}{2\lambda_\Lambda}
\end{align}
in the chiral limit.  The Lagrangian density \eq{eq:MFlag} bilinear in
the quark fields, and hence the path integral quantization is done by
a Gaussian integration. The resulting thermodynamic potential is 
a function of $M$ and is given by 
\begin{align}\label{eq:Om/V}
 \beta\Omega/V &= -2\Nc\Nf\int_{p^2\leq \Lambda^2}\frac{\rmd^3 p}{(2\pi)^3}
  \Bigl\{ \omega + T\ln\bigl[1\!+\!\rme^{-\beta(\omega-\muq)}\bigr] \notag\\
 &\qquad +T\ln\bigl[1\!+\!\rme^{-\beta(\omega+\muq)}\bigr] \Bigr\}
  +\frac{M^2}{2\lambda_\Lambda}
\end{align}
with the quasi-particle energy dispersion relation,
$\omega=\sqrt{p^2+M^2}$.  The momentum integral is cut at the cutoff
scale as the mean field Lagrangian~\eqref{eq:MFlag} supposedly arising
from integrating-out the quarks with momenta larger than the cutoff
scale, $p^2>\Lambda^2$.  The overall constant $2$ comes from the spin
degeneracy.  For the moment we restrict ourselves to vanishing
temperature and density; $T=\muq=0$.  Later we extend the analysis to
include thermal and medium effects.  In this simple setup the momentum
integral in Eq.~\eqref{eq:Om/V} can only lead to a function of
$\xi=M/\Lambda$ only, up to a dimensional factor $\Lambda^4$.  The
zero-point energy contribution gives the leading contribution to the
thermodynamic potential in Eq.~\eq{eq:Om/V} and can be evaluated
explicitly as, e.g.~\cite{Hatsuda:1994pi},
\begin{align}
 &2\Nc\Nf\int_{p^2\leq \Lambda^2}\frac{\rmd^3 p}{(2\pi)^3}\,\omega \notag\\
 &= \frac{\Nc\Nf\Lambda^4}{8\pi^2}\Biggl[
  \sqrt{1+\xi^2}\,(2+\xi^2) + \frac{\xi^4}{2}\ln\biggl|
  \frac{\sqrt{1+\xi^2}-1}{\sqrt{1+\xi^2}+1}\biggr| \Biggr] \notag\\
 &\simeq \frac{\Nc\Nf\Lambda^4}{4\pi^2} \bigl( 1+\xi^2 \bigr) +O(\xi^2)\;. 
\label{eq:xi-expansion}
\end{align}
Note that we do not loose the generality of our analysis by assuming
small $\xi=M/\Lambda$ because we are only interested in the onset of
chiral phase transition where $\xi$ starts taking a finite value from
zero.  With the expansion in $\xi$ the thermodynamic potential
simplifies, to wit,
\begin{equation}\label{eq:Osimple}
 \beta\Omega/V \simeq -\frac{\Nc\Nf\Lambda^4}{4\pi^2}
  \biggl[ 1+\biggl( 1-\frac{2\pi^2}{\Nc\Nf\, \lambda_\Lambda
  \Lambda^2} \biggr)\xi^2 \biggr] \;, 
\end{equation}
where we have dropped the higher order terms in $\xi$.  The critical
coupling associated with the second-order phase transition should
correspond to a point at which the curvature of the thermodynamic
potential crosses zero.  In order to have a non-zero $\xi$ the
potential curvature should be negative and this condition leads to an
inequality~\cite{Hatsuda:1994pi},
\begin{equation}
 \lambda_\Lambda \Lambda^2 > \frac{2\pi^2}{\Nc\Nf} \;.
\label{eq:critical}
\end{equation}
This is the condition for the NJL model to accommodate for spontaneous
chiral symmetry breaking.  Seemingly this depends on the UV cutoff
scale $\Lambda$.  However, as we have already argued above, $\Lambda$
relates to $\Lambda_{\rm QCD}$ in the current effective field theory
set-up.

Finite temperature effects are taken into account in a straightforward
extension.  The full dependence including the finite-$\muq$ effect
will be considered in the numerical analysis in Sec.~\ref{sec:RG}.
Again we aim at simplicity and analytically estimate the finite-$T$
corrections.  This leads to simple understanding of how the chiral
phase transition takes place within this effective model.  Since we
are interested in the regime with infinitesimal $\xi$, we can make use
of the high-$T$ expansion for $T\gg M$, which yields the following
thermodynamic potential,
\begin{equation}
 \beta\Omega/V \simeq -\frac{\Nc\Nf\Lambda^4}{4\pi^2} \Biggl\{ 1
  + \biggl[ 1 - \frac{2\pi^2}{\Nc\Nf\, \lambda_\Lambda \Lambda^2}
  -\frac{\pi^2}{3}\biggl(\frac{\Lambda}{T}\biggr)^2 \biggr]\xi^2
   \Biggr\} \;.
\end{equation}
Then, in the same way as the condition~\eqref{eq:critical} has been
derived, we see that spontaneous chiral symmetry breaking requires,
\begin{equation}
 \lambda_\Lambda \Lambda^2 > \frac{2\pi^2}{\Nc\Nf}
  \Biggl[ 1-\frac{\pi^2}{3}\biggl(\frac{\Lambda}{T}\biggr)^2 \xi^2
  \Biggr] \;.
\label{eq:critical-t}
\end{equation}
In the effective model studies, usually, the four-Fermi coupling is
fixed by observables in the vacuum.  For a given $\lambda_\Lambda$
that satisfies the above inequality, we can read the critical
temperature directly from Eq.~\eqref{eq:critical-t} as
\begin{equation}
 \Tc = \sqrt{ \frac{3\Lambda^2}{\pi^2}
  - \frac{6}{\Nc\Nf \lambda_\Lambda} } \;.
\label{eq:Tc-mf-zeroB}
\end{equation}
We readily confirm that Eq.~\eqref{eq:Tc-mf-zeroB} gives a
conventional value for the critical temperature, i.e.\ $\Tc=175\MeV$
by plugging-in the standard NJL-model parameters,
e.g.~\cite{Hatsuda:1994pi}: $\Lambda=631\MeV$ and
$\lambda_\Lambda/2=0.214\fm^2$.

\subsection{Effects of the magnetic field}

We proceed by introducing an external magnetic field $\bB$.  In the
present work we restrict ourselves to spatially homogeneous and
temporally constant $\bB$ fields.  In the presence of $\bB$ (along the
$z$-axis) the transverse momenta (i.e.\ the $x$ and $y$ components) of
spin-$1/2$ particles with electric charge $q$ are quantized into the
Landau levels and the momentum integration with spin sum of a general
function $f(\omega)$ is replaced as
\begin{equation}
 \begin{split}
 &2\int_{p^2\leq \Lambda^2} \frac{\rmd^3 p}{(2\pi)^3}\,f(\omega)
  \;\longrightarrow\;
  2\int_{\Lambda,B} \frac{\rmd^3 p}{(2\pi)^3}\,f(\omega_n) \\
 &\di\equiv \frac{|qB|}{2\pi}\sum_{n=0}^{N_{\Lambda,B}} \alpha_n
  \int_{p^2\leq \Lambda^2 }
  \frac{\rmd p_z}{2\pi}\, f\bigl(\sqrt{p_z^2 + 2|qB|n +M^2}\bigr) \;,
 \end{split}
\label{eq:Landau}
\end{equation}
where $2$ in the left-hand side is the spin degeneracy, which is
replaced by the spin-degeneracy factor $\alpha_n$ that is $1$ for
$n=0$ (Landau zero-mode) and $2$ otherwise.  In the second line,
$p^2=p_z^2+2|qB|n$ and $N_{\Lambda,B}=\theta_g[\Lambda^2/(2|qB|)]$,
where $\theta_g[n+x]=n$ for $n\in \mathbb{Z}$ and $0<x<1$. 

Now, let us continue our analytical consideration taking the simplest
limit of large magnetic field, more specifically
$2|qB|> \Lambda^2$.  In this particular limit only the Landau
zero-mode contributes as $\theta_g[\Lambda^2/(2|qB|)]=0$ and the
lowest Landau level approximation (LLLA) is exact.  Note that this
should be already a rather quantitative approximation for heavy-ion
collisions as the magnetic field strength there is of order of
$\Lambda_{\rm QCD}$ which is related to our cutoff scale $\Lambda$.
In any case we shall also consider general magnetic field within our
numerical calculations.  In the LLLA, at $\muq=0$, we can express the
thermodynamic potential as
\begin{equation}
 \begin{split}
 \beta\Omega/V &= -\Nc \sum_{f=u,d} \frac{|q_f B|}{2\pi}
  \int\frac{\rmd p_z}{2\pi} \biggl[ \omega_0 \\
 &\qquad\qquad\qquad +2T\ln\bigl( 1+\rme^{-\beta \omega_0} \bigr)\biggr]
  +\frac{M^2}{2\lambda_\Lambda} \;.
 \end{split}
\end{equation}
Here we defined $\omega_0=\sqrt{p_z^2+M^2}$ and the quark electric
charges are $q_u=(2/3)e$ and $q_d=-(1/3)e$.  From this form of the
thermodynamic potential one can easily understand the essence of the
Magnetic Catalysis.  At $T=0$ we can expand the zero-point
energy for small $\xi=M/\Lambda$, as we did previously, to find the
thermodynamic potential up to $\mathcal{O}(\xi^2)$ as
\begin{equation}
 \beta\Omega/V \simeq -\Nc\sum_{f=u,d} \frac{|q_f B|\Lambda^2}{4\pi^2}
  \biggl[ 1+\Bigl(\ln\frac{2}{\xi} + \frac{1}{2}\Bigr)\xi^2
  \biggr] + \frac{\Lambda^2}{2\lambda_\Lambda}\xi^2 \;.
\label{eq:B-mf-potential}
\end{equation}
As we have already argued, the sign of the curvature determines
whether chiral symmetry is spontaneously broken or not.  The
remarkable feature in the LLLA is that $\ln(1/\xi)$ in the coefficient
of $\xi^2$ is positive and arbitrarily large for $\xi\to0$ and thus it
can eventually overcome $\Lambda^2/(2\lambda_\Lambda)$ from the last
term.  This means that chiral symmetry is always broken however small
$\lambda_\Lambda$ is.  The important point is that such a logarithmic
term appeared in the coefficient of $\xi^4$ in
Eq.~\eqref{eq:xi-expansion} when $\bB$ was not applied.  In the LLLA
with strong $\bB$ the phase-space volume is suppressed and $\xi^4$ is
reduced to $(|q_f B|/\Lambda^2)\xi^2$.  Because $\ln(1/\xi)$ changes
its sign from positive to negative for large $\xi$, the thermodynamic
potential~\eqref{eq:B-mf-potential} has a minimum as a function of
$\xi$ that solves the gap equation.  In fact, by taking the
derivative, we can write down the gap equation,
$-\Nc \sum_f (|q_f B|\Lambda^2 / 2\pi^2)\ln(2/\xi)
 +\Lambda^2/\lambda_\Lambda = 0$, whose solution $\xi_0$ or the
corresponding constituent mass is
\begin{equation}
 M_0 = \xi_0\Lambda = 2\Lambda
  \exp\biggl( -\frac{2\pi^2}{\Nc\,\lambda_\Lambda \sum_f|q_f B|}
  \biggr) \;.
\label{eq:mag_mass}
\end{equation}
This expression for the condensate is quite analogous to the
superconducting gap in the BCS theory.  We can see that the critical
temperature at which the condensate melts is also given by the same
relation as the one known in the BCS theory.

We can carry out the high-$T$ expansion in the same way as previously
using
\begin{equation}
 \int_{-\infty}^\infty\frac{\rmd p_z}{2\pi} \ln(1+\rme^{-\beta\omega_0})
 \simeq \frac{\pi T}{12} + \frac{\Lambda^2}{4\pi T} \Bigl(
 \ln\frac{\Lambda \xi}{\pi T} + \gamma - \frac{1}{2} \Bigr)\xi^2 \;,
\end{equation}
up to $\mathcal{O}(\xi^2)$.  Interestingly enough, a logarithmic term
proportional to $\xi^2\ln(1/\xi)$ appears from this finite-$T$
integration as well as the vacuum contribution that we already
mentioned.  We can make sure that these logarithmic singularities
exactly cancel out and the Magnetic Catalysis is lost at finite $T$.
This trend had been reported in the literature~\cite{Suganuma:1990nn}
and was found also in the context of the Magnetic
Catalysis~\cite{Das:1995bn}.  In the RG analysis in Sec.~\ref{sec:RG}
the cancellation of the IR singularities will be reconfirmed in a more
transparent manner.

The curvature of the thermodynamic potential no longer has a
logarithmic singularity at finite $T$ and the critical coupling
constant becomes finite again.  This is actually the reason why we can
expect a chiral phase transition at a certain temperature even in
the LLLA calculation with strong $\bB$.  After all, the thermodynamic
potential~\eqref{eq:B-mf-potential} receives the finite-$T$
corrections as
\begin{equation}
 \begin{split}
 \beta\Omega/V &\simeq -\Nc\sum_f \frac{|q_f B|\Lambda^2}{4\pi^2}
  \Biggl\{ 1+\frac{\pi^2}{3}\biggl(\frac{T}{\Lambda}\biggr)^2 \\
 &\quad\;\; + \biggl[ \ln\biggl(\frac{2\rme^\gamma \Lambda}{\pi T}\biggr)
  - \frac{2\pi^2}{\Nc \lambda_\Lambda\sum_f|q_f B|} \biggr]\xi^2
  \Biggr\} \;.
 \end{split}
\end{equation}
The zero of the coefficient of $\xi^2$ gives a condition for the
critical coupling or the critical temperature as obtained in
Eq.~\eqref{eq:Tc-mf-zeroB}.  In this case of the LLLA we arrive at
\begin{equation}
 \Tc = \frac{2\rme^\gamma \Lambda}{\pi}
  \exp\biggl( -\frac{2\pi^2}{\Nc\,\lambda_\Lambda
  \sum_f|q_f B|} \biggr) = \frac{\rme^\gamma}{\pi} M_0 \;.
\label{eq:Tc-mf-B}
\end{equation}
This relation between the condensate at $T=0$ and the melting
temperature is perfectly identical to the results from the BCS theory.
If the coupling constant $\lambda_\Lambda$ is infinitely large, all
expressions appear very simple; the critical temperature without $\bB$
is read from Eq.~\eqref{eq:Tc-mf-zeroB} as
$\Tc=(\sqrt{3}/\pi)\Lambda\simeq 0.551\Lambda$, and that at strong
$\bB$ is from Eq.~\eqref{eq:Tc-mf-B} as
$\Tc=(2\rme^\gamma/\pi)\Lambda\simeq 1.13\Lambda$.  Therefore the
critical temperature is indeed raised by the $\bB$ effect, which is in
qualitative agreement with effective model calculations.

\section{Renormalization Group Analysis}
\label{sec:RG}

\begin{figure*}
\begin{center}
 \includegraphics[width=0.8\textwidth]{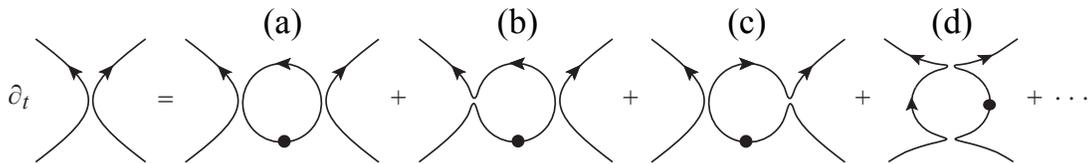}
\end{center}
 \caption{Diagrammatic representation of the flow equation for the
   four-Fermi interaction.  At each vertex the color index is
   contracted along the connected lines.  The ellipses are neglected
   contributions from higher loop diagrams beyond the
     Ansatz~\eqref{eq:Ansatz}}.
 \label{fig:flow}
\end{figure*}

In the previous section we have discussed the standard mean-field
analysis where quark quantum fluctuations are taken into account at
one loop.  In the present section we extend the analysis beyond the
one-loop level by means of RG-techniques.  This also allows us to
formulate the physics mechanism of the chiral phase transition from a
different and rather simple point of view.  To that end we here
introduce an IR-cutoff term into the Lagrangian and derive a
functional RG equation for the free energy or effective action, the
Wetterich equation~\cite{Wetterich:1992yh}.  In this approach the
kinetic term is modified in momentum space in order to suppress
momentum modes below the given infrared (IR) cutoff scale $k$.  This
leads to an infrared regularized free energy or effective action
$\Gamma_k[\psi,\bar\psi]$ which reduces to the thermodynamic potential
$\Omega$ at vanishing IR cutoff $k=0$ on the equations of motion
(EoM) of $\psi,\bar\psi$, that is $\Omega= \Gamma_{k=0}|_{\rm EoM}$.
For the sake of simplicity we resort to a three-dimensional IR-cutoff
which allows us to perform the Matsubara summation analytically.  Then
the fermion propagator in Euclidean space takes the following form,
\begin{equation}
 G(p) = \frac{1}{\gamma_4 p_4 + \bgamma\cdot\bp\, [1+r_k(\bp)]} \;.
\end{equation}
Here $r_k(\bp)$ is the cutoff function with $r_k(\bp)\sim
k/|\bp|$ for $\vec p^2<k^2$ and $r_k(\bp)\to 0$ for $\bp^2>k^2$.  Note
also that the dimensionless shape of $r_k$ is genuinely a function of
$|\bp|/k$.  Since we are interested in the chiral properties of QCD,
we have avoided implementing the cutoff function as a
momentum-dependent mass term but in the kinetic term.  The former
cutoff explicitly breaks chiral symmetry and would have similar
effects on chiral properties as the Wilson mass term has on the
lattice.  The scale-dependence of $\Gamma_k$ is then encoded in the
flow equation, schematically written as 
\begin{align}\nonumber
 \partial_t \Gamma_k[\psi,\bar\psi] &= -\Tr\,\Biggl(
  \0{1}{\Gamma_k^{(1,1)}[\psi,\bar\psi] +\bgamma\cdot\bp\;r_k}
  \,\bgamma\cdot\bp\; \partial_t r_k \Biggr)\;,\\[1ex]
 \Gamma_k^{(1,1)}[\psi,\bar\psi] &= \0{\overrightarrow{\delta}}{
  \delta\bar{\psi}} \Gamma_k[\psi,\bar\psi]  
  \0{\overleftarrow{\delta}}{ \delta{\psi} }\;, 
\label{eq:flowGA} 
\end{align}
where $t=\ln(k/\Lambda)$ is the logarithmic IR scale with reference
scale $\Lambda$ and the trace $\Tr$ sums over momenta, Dirac indices,
flavors.  (For QCD-related reviews, see
Refs.~\cite{Litim:1998nf,Berges:2000ew,Pawlowski:2005xe,Gies:2006wv,%
Braun:2011pp}).  In the present work we ignore the effects of the
multi-scattering of quarks, which is phase-space suppressed.  We rush
to add that this argument gets weaker at higher densities.  For the
time being we make the additional approximation of a local effective
four-Fermi vertex $\lambda_k$, which will be relaxed in the next
section.  This leads us to the following simple Ansatz for the
effective action,
\begin{equation}
 \Gamma_k = \int\rmd^4 x \biggl(\bar{\psi}\rmi\feyn{\partial}\psi
  +\frac{1}{2}\bar{\psi}^{a\alpha}_i \psi^{b\alpha}_j
  \Gamma_k{}_{ijlm}^{abcd} \bar{\psi}^{c\beta}_l \psi^{d\beta}_m
  \biggr) \;, 
\label{eq:Ansatz}
\end{equation}
where $\Gamma_k{}_{ijlm}^{abcd}$ is given by Eq.~\eqref{eq:Gamma} with
$\Lambda\to k$.  This form is substituted into the flow
equation~\eqref{eq:flowGA} and we take the functional derivatives.
The left-hand side of the flow equation is then,
\begin{align}
 & \partial_t\,\Gamma_k^{(2,2)} \equiv \partial_t\,
 \frac{\overrightarrow{\delta}}{\delta\bar{\psi}_i^{a\alpha}(x)}
 \frac{\overrightarrow{\delta}}{\delta\bar{\psi}_l^{c\gamma}(z)}
 \Gamma_k
 \frac{\overleftarrow{\delta}}{\delta\psi_j^{b\beta}(y)}
 \frac{\overleftarrow{\delta}}{\delta\psi_m^{d\delta}(w)} \notag\\
 & = \delta(x-y)\,\delta(y-z)\,\delta(z-w) \notag\\
 &\qquad \times \bigl(\delta^{\alpha\beta}\delta^{\gamma\delta}
 \partial_t\Gamma_k{}_{ijlm}^{abcd} -\delta^{\alpha\delta}\delta^{\beta\gamma}
 \partial_t\Gamma_{k}{}_{imlj}^{adcb} \bigr) \;.
\label{eq:deriv}
\end{align}
The right-hand side is a bit complicated.  There arise twelve terms,
six of which are proportional to
$\delta^{\alpha\beta}\delta^{\gamma\delta}$ and other six proportional
to $\delta^{\alpha\delta}\delta^{\beta\gamma}$, respectively.  Here,
let us focus on the part of
$\delta^{\alpha\beta}\delta^{\gamma\delta}$ only.  In addition,
because we aim to have the flow equation for $\lambda_k$, we can
simplify the calculation by contracting the Dirac and the flavor
indices of external legs, namely,
$\delta_{ij}\delta_{lm}\delta^{ab}\delta^{cd}\Gamma_{k}{}_{ijlm}^{abcd}
=16\Nf^2\, \lambda_k$.  Then six terms are no longer independent but
only three remain distinguishable.  The flow equation is eventually
given by
\begin{align}
 &\partial_t \Gamma_{k}{}_{iill}^{aacc}
  =2\Nc\Tr\bigl[ G_{jm}\Gamma_{mpii}^{bdaa}G_{pq}\Gamma_{qrll}^{dbcc}
  G_{rs}(\bgamma\cdot\bp)_{sj} \;\partial_t r_k \bigr] \notag\\
 &\qquad -2\Tr\bigl[ G_{jm}\Gamma_{miip}^{baad}G_{pq}
  \Gamma_{qrll}^{dbcc}G_{rs}(\bgamma\cdot\bp)_{sj}
  \;\partial_t r_k \bigr] \notag\\
 &\qquad -2\Tr\bigl[ G_{jm}\Gamma_{mpii}^{bdaa} G_{pq}
  \Gamma_{qllr}^{dccb} G_{rs} (\bgamma\cdot\bp)_{sj} \; \partial_t
  r_k \bigr] \;,
\label{eq:flow}
\end{align}
when no momentum is carried by any external legs.  In
Eq.~\eqref{eq:flow} we have also dropped the subscript $k$ on the
vertices.  There are four different types of Feynman diagrams that
contribute to the flow equation as displayed in Fig.~\ref{fig:flow}.
Here (a) corresponds to the first term in the right-hand side of
Eq.~\eqref{eq:flow}, (b) and (c) to the second term, and (d) to the
third term.

After some calculations we can arrive at the following form of the
flow equation with $\Nf=2$;
\begin{align}
   &\partial_t \lambda_k = -4 ~\left(2\Nc+\0{3}{4}\right)
   \lambda_k^2\sum_{f=u,d}\,\, \sumint_{T,B}
   \frac{\rmd^4 p}{(2\pi)^4} \,|\bp| \partial_t r_k(\bp)\, \notag\\
   &\qquad\qquad\qquad\qquad\times \frac{ |\bp|\,
     [1+r_k(\bp)]}{\bigl\{p_4^2+\bp^2[1+r_k(\bp)]^2\bigr\}^2} \;,
\label{eq:dtlambda}
\end{align}
where the factor four in Eq.~\eqref{eq:dtlambda} originates in the
trace of the Dirac index and the factor $2\Nc$ comes from (a)
  which is the leading diagram in the $1/\Nc$ expansion.  The factor
$3/4=1-1/4$ is a combination of 1 from the diagrams (b) and (c)
and $1/4$ from the diagram (d).

Before we come to the explicit computations we would like to discuss
the structure of the flow in Eq.~\eqref{eq:dtlambda} as it already
reveals a lot of interesting properties of the flow and is also
important for the computations and the interpretation of the results.
The flow~\eqref{eq:dtlambda} has a total $t$-derivative structure and
hence leads to results at $k=0$ that are independent of the chosen
regulator $r_k$ up to (re-)normalizations at the initial scale
$\Lambda$.  To see this more clearly we rewrite it as a flow for
$1/\lambda_k$, to wit,
\begin{equation}
 \partial_t \0{1}{\lambda_k} = -\partial_t \bigl[\Pi_k(0)
  -\Pi_\Lambda(0)\bigr]\;,
\label{eq:dt1ovlambda}
\end{equation}
with 
\begin{equation}
 \Pi_k(0) \simeq  - \sumint_{T,B}  \frac{\rmd^4 p}{(2\pi)^4}
  \sum_{f=u,d} \frac{2 \bigl(2\Nc+\0{3}{4}\bigr)}
  {p_4^2+\bp^2\,[1+r_k(\bp)]^2} \;.
\label{eq:Pik}
\end{equation} 
In Eq.~\eqref{eq:dt1ovlambda} we have introduced the subtraction term
for having explicitly finite expressions before $t$-derivatives are
taken.  In its form~\eqref{eq:dt1ovlambda} the flow is easily
integrated and leads to
\begin{equation} \label{eq:resultlambda}
 \0{1}{\lambda_k}= \0{1}{\lambda_\Lambda} - \bigl[\Pi_k(0)
  -\Pi_\Lambda(0)\bigr]\,.
\end{equation}
We immediately deduce that the dependence on the regulator drops out
at $k=0$ for consistent choice of $\lambda_\Lambda$;  evidently there
is no dependence on $r_0(p)\equiv 0$.  In turn,
$1/\lambda_\Lambda+\Pi_\Lambda(0)$ is required to be
$\Lambda$-independent, otherwise $\lambda_0$ would depend on
$\Lambda$.  The latter condition encodes RG-invariance of the theory,
not to be mixed-up with renormalizability.  We also deduce from
Eq.~\eqref{eq:resultlambda} that in the leading order
$1/\lambda_\Lambda\propto -\Pi_\Lambda(0) =c(r_\Lambda)\Lambda^{2}$.
The prefactor depends on the chosen regulator and hence defines the
regularization scheme.  This is seen more clearly if comparing the
dimensionless coupling,
\begin{equation}\label{eq:hatlambda}
 \hat\lambda_k = \lambda_k k^2 \;,
\end{equation}
at the initial scale $\Lambda$ for two different regulator
$r^{(1)}_k,r^{(2)}_k$.  This leads to $\hat \lambda_\Lambda^{(1)}/
\hat\lambda_\Lambda^{(2)}=c^{(2)}/c^{(1)}$.  For agreeing physics
scales the above ratio should be one and we identify
$\Lambda^{(2)}=c^{(1)}/c^{(2)}\Lambda$ with $\Lambda^{(1)}=\Lambda$.
In other words, we expect that the equivalence of the mean-field
inequality~\eq{eq:critical} depends on the chosen regulator simply by 
a multiplicative factor in front of $\hat\lambda$;
\begin{equation}
 \lambda_\Lambda \Lambda^2 >
 \0{c(r_\Lambda)}{c(r^{\rm sharp}_\Lambda)}\frac{2\pi^2}{\Nc\Nf} \;.
\label{eq:criticalRG}
\end{equation}
where $c(r^{\rm sharp}_\Lambda)$ stands for the implicit regulator
used in the mean-field computation, that is the three-dimensional
sharp cutoff; $r^{\rm sharp}_k(p^2)=\theta(p^2-k^2)$.  We emphasize
once more that this only reflects the dependence of $\lambda_\Lambda$
on the chosen RG-scheme.

In summary we conclude that $\lambda_0$ does not depend on $r_\Lambda$
either and any regulator-dependence is removed.  We emphasize that
this is a particularity of the present approximation and hinges on its
explicit total derivative structure.  Such approximations are
singled-out as regulator-independent and also related to optimization
criteria that amongst other properties (re-)enforce total derivative
structures for the flows; for more detailed discussions, see
Ref.~\cite{Pawlowski:2005xe,Blaizot:2010zx}.  For later purpose we
also resolve Eq.~\eqref{eq:resultlambda} for $\lambda_k$,
\begin{equation}\label{eq:resultlambdaex}
 \lambda_k = \0{\lambda_\Lambda}
  {1-\lambda_\Lambda[\Pi_k(0)-\Pi_\Lambda(0)]}\;.
\end{equation}
Equation~\eqref{eq:resultlambda} elucidates the resummation structure
of the flow.  It encodes a bubble resummation of the diagrams in
Fig.~\ref{fig:flow}.  This is seen within a diagrammatic expansion of
Eq.~\eqref{eq:resultlambda} in orders of $\lambda_\Lambda$,
\begin{equation}
 \includegraphics[width=0.8\columnwidth]{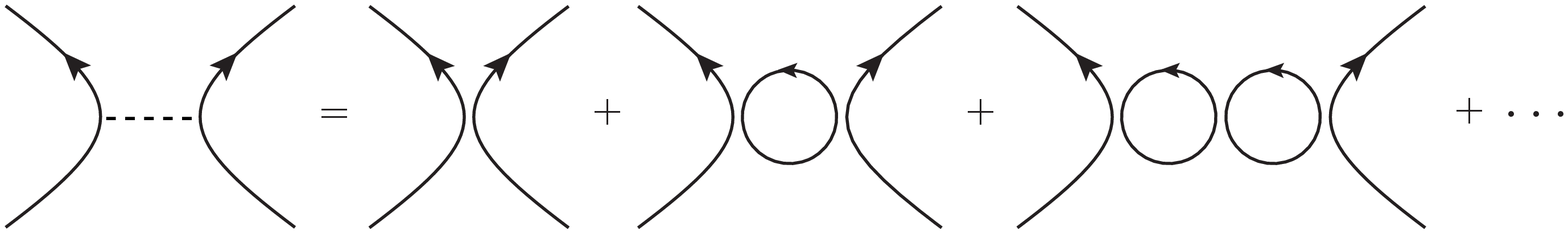}
\label{eq:schematic}
\end{equation}
If we interpret the dashed line on the left-hand side as a meson
propagator it results simply from the bubble resummation shown in
Fig.~\ref{eq:schematic}.  The first term on the right-hand side, i.e.\
the tree-level diagram, gives $\lambda_\Lambda$ and hence a tree-level
meson propagator $\lambda_\Lambda$.  The second term comes from the
one-loop polarization, that is, $\lambda_\Lambda^2 \Pi_k(0)$, where
the momentum argument is zero because we only consider the situation
that all external legs bring no momentum insertion.  If we lower the
temperature towards the hadronic phase the four-Fermi vertex function
has to diverge at the phase transition as it contains a massless sigma
and pion propagation.  Indeed, when introducing effective mesonic
degrees of freedom with the $\sigma$ and $\vec \pi$ coupling to the
related quark bilinears, it can be shown that the meson mass is
proportional to $1/\lambda_k$.  This makes it clear that a single pole
in the four-Fermi coupling does not merely signal the validity bound
of the approximation but rather the onset of condensation.

This ends our discussion of the formal properties of the
flow~\eqref{eq:dtlambda} and we proceed on how to practically solve
it.  As we have argued, its solution does not depend on the chosen
regulator.  This fully justifies the choice taken already in
restricting ourselves to three-dimensional regulators.  It allows us
to perform the $p_4$ integration or summation with respect to the
Matsubara frequency most easily.  The result is
\begin{equation}\label{eq:Matsubara}
 T\sum_n \frac{\omega}{(p_4^2+\omega^2)^2} =
  \frac{1}{4\omega^2} +
  \frac{\rmd}{\rmd\omega}\frac{1}{2\omega} \frac{1}{\rme^{\beta\omega} + 1} \;,
\end{equation}
with $\omega\equiv|\bp|\,|1+r_k(\bp)|$.  We still have to specify the
three-dimensional regulator, $r_k(p)$, to proceed further.  We again
utilize the regulator-independence and adopt the three-dimensional
flat cutoff, \cite{Litim:2006ag}, for the sake of simplicity,
\begin{equation}
 r^{\rm flat}_k(\bp) = \biggl(\frac{k}{|\bp|}-1\biggr)\,
  \theta(k/|\bp|-1) \;. 
\label{eq:rk}
\end{equation}
Such flat cutoffs were first introduced in Ref.~\cite{Litim:2000ci}
for the purpose of optimization are also most useful for analytical
studies.  It is called ``flat'' because it renders the spatial
dispersion momentum-independent for momenta below the cutoff scale,
\begin{equation}\label{eq:flat}
 \omega(r_k^{\rm flat}) \theta(k/|\bp|-1)= k \;, 
\end{equation}
where $\omega$ denotes a quantity defined just blow Eq.~\eqref{eq:Matsubara}.
For momenta $|\bp|<k$, hence, the Matsubara sum in
Eq.~\eqref{eq:Matsubara} turns into
\begin{equation}\label{eq:flatMatsubara} 
 \0{1}{2 k^2}\biggl[\012 -\0{1}{\rme^{\beta k}+1}
 -\beta k\0{\rme^{\beta k}} {(\rme^{\beta k}+1)^2}\biggr] \;.
\end{equation}
which is completely momentum-independent.  The integrand in the
four-Fermi flow~\eqref{eq:dtlambda} is also proportional to the scale
derivative of the regulator,
$\partial_t\, r^{\rm flat}_k(p) = (k/|\bp|)\,\theta(k/|\bp|-1)$ which
limits the integration to momenta $|\bp|<k$ where the integration is
trivial.  This leads us, particularly when $\bB=0$, to
\begin{equation}\label{eq:dtPiexplicit} 
 \partial_t  \Pi_k(0)=
 \0{\bigl(2\Nc+\0{3}{4}\bigr)\Nf }{3 \pi^2}
 \biggl[\012 -\0{1}{\rme^{\beta k}+1}
 -\beta k\0{\rme^{\beta k}}{(\rme^{\beta k}+1)^2}\biggr] \;,
\end{equation}
to be inserted in Eq.~\eqref{eq:dt1ovlambda}.  In later discussions it
also turns out that the choice~\eqref{eq:flat} for $r_k(\bp)$ avoids
unphysical oscillatory behavior in the flow from the discrete Landau
levels.  For going beyond the present approximation which we shall do
in the last section it is worth noticing that this cutoff is optimized
in the high temperature limit, \cite{Litim:2000ci,Pawlowski:2005xe},
and keeps some of this property also at momentum-independent
approximations.  For fully momentum-dependent approximations as well
as finite magnetic field it introduces momentum non-localities
(momentum flows in diagrams) which affect the quantitative accuracy
(see Ref.~\cite{Fister:2011uw}).  In short, within the approximations
used in the present work the regulator works sufficiently well and has
the advantage of analytical tractability.  Within extensions of this
work we shall aim at quantitative precision and will resort to
regulators which maintain momentum locality and minimize the momentum
flow.

\subsection{Flow and chiral symmetry breaking}

We first concentrate on the relation between the flow of the
four-Fermi coupling and chiral symmetry breaking at $T=0$ and
$\bB=0$.  This gives access to the simple structure and understanding
of chiral symmetry breaking within the flow equation approach, and
hence serves as a useful example for understanding the results in the
presence of medium effects.  In order to facilitate the direct
comparison with the mean-field results, we shall first discuss the
large-$\Nc$ limit at leading order.  Then, only the diagram (a) in
Fig.~\ref{fig:flow} contributes.  Note that due to the simple full
$N_c$-dependence of the flow (see Eq.~\eqref{eq:dtlambda}), the limit
is easily undone by replacing $\Nc\to\Nc+\frac{3}{8}$ in the large
$\Nc$ result.  For the flat regulator and $T=0$ the flow of the vacuum
polarization, Eq.~\eqref{eq:dtPiexplicit}, is $\Nc/(3 \pi^2)$ and we
arrive at
\begin{equation}
 \partial_t \lambda_k = -\frac{\Nc\Nf\, \lambda_k^2 k^2}{3\pi^2} \;, 
\label{eq:flow_zeroTB}
\end{equation}
as well as
\begin{equation}
 \lambda_k = \frac{\lambda_\Lambda}{\displaystyle
  1+\frac{\Nc\Nf\, \lambda_\Lambda}{6\pi^2}(k^2-\Lambda^2)} \;,
\label{eq:running_zeroTB}
\end{equation}
with a positive initial coupling $\lambda_\Lambda$.  From
Eqs.~\eqref{eq:flow_zeroTB} and \eqref{eq:running_zeroTB} we infer
that lowering $k$ increases $\lambda_k$.  At
$\Nc\Nf\, \lambda_\Lambda \Lambda^2/(6\pi^2) = 1$ the coupling
diverges and has a single pole.  We have already argued that
Eq.~\eqref{eq:criticalRG} has to be understood as the onset of
spontaneous chiral symmetry breaking, and the occurance of massless
mesonic modes.  Hence, with the present regulator the condition for
spontaneous chiral symmetry breaking, Eq.~\eqref{eq:criticalRG}, is
given by
\begin{equation}
 \lambda_\Lambda \Lambda^2 > \frac{6\pi^2}{\Nc\Nf} \;.
\label{eq:critical_rgflat}
\end{equation}
This result is the counterpart of the condition~\eq{eq:critical} in
the mean-field theory.  The discrepancy by a factor $3$ relates to the
different regularization scheme between the sharp cutoff in the
mean-field treatment and the optimized cutoff in the RG analysis.  We
can identify, roughly speaking, the RG-scale $\Lambda^2$ in the
present RG study with $3\Lambda^2$ in the mean-field approximation.

The same conclusion can be drawn from a fixed-point argument, directly
utilizing Eq.~\eqref{eq:flow_zeroTB};  by changing to the
dimensionless coupling $\hlambda$ in Eq.~\eqref{eq:hatlambda} the
flow~\eqref{eq:flow_zeroTB} changes to
\begin{equation}
 \beta_{\hlambda} \equiv \partial_t \hlambda_k
  = 2\hlambda_k - \frac{\Nc\Nf}{3\pi^2}\, \hlambda_k^2 \;.
\label{eq:beta}
\end{equation}
The $\beta$-function in Eq.~\eqref{eq:beta} vanishes at
$\hlambda_\ast = 6\pi^2/(\Nc\Nf)$ which is an attractive UV fixed
point.  Of course, this defines exactly the critical coupling constant
given in Eq.~\eqref{eq:critical_rgflat}.  If $\hlambda_\Lambda$ starts
from a value above $\hlambda_\ast$, the flow grows larger for smaller
$k$, which indicates the spontaneous breaking of chiral symmetry.

\begin{figure}
\begin{center}
 \includegraphics[width=0.8\columnwidth]{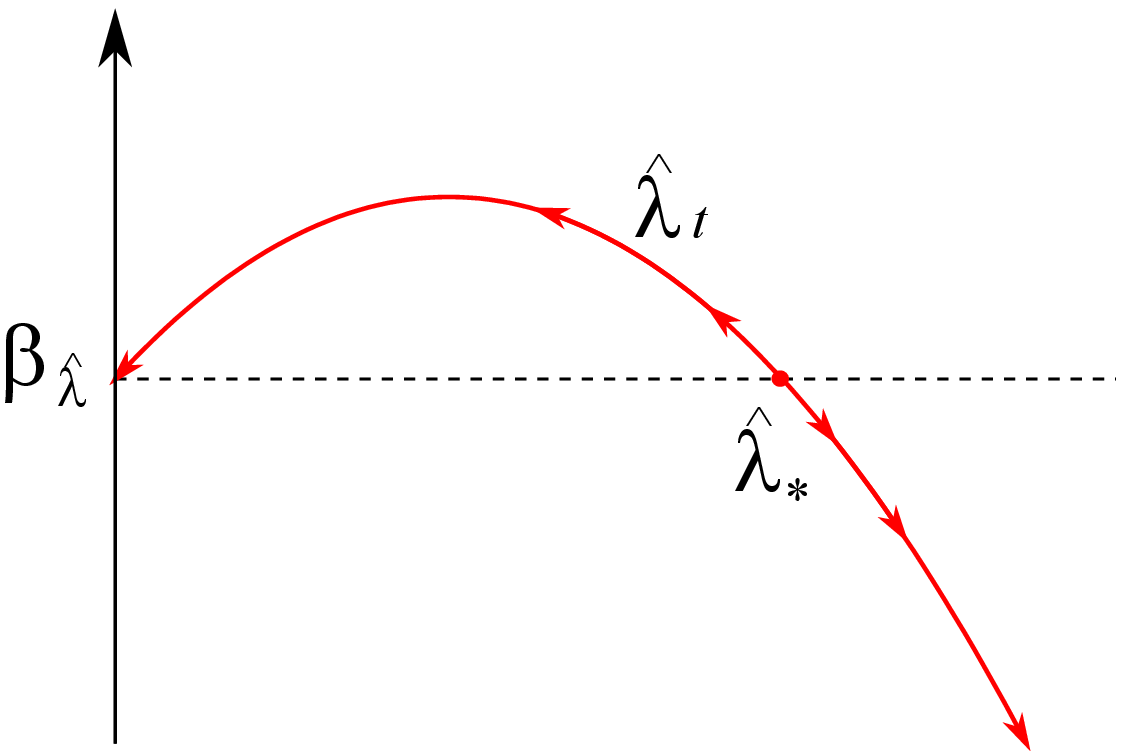} \vspace{3mm}\\
 \includegraphics[width=0.8\columnwidth]{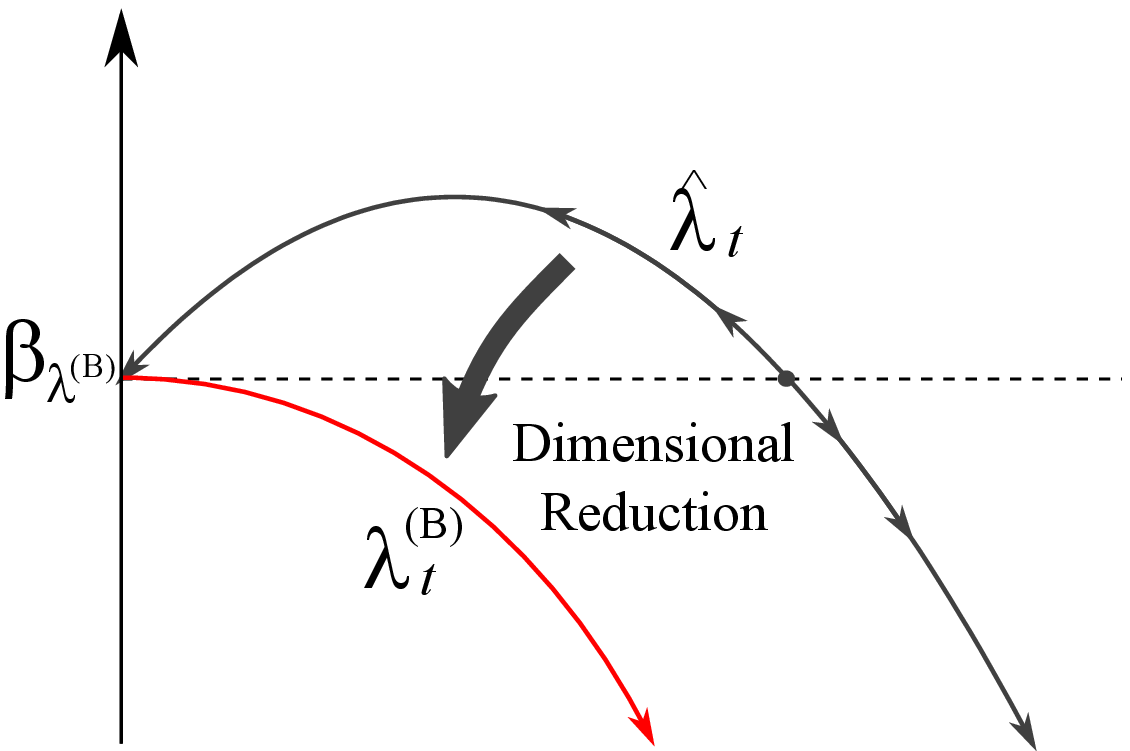}
\end{center}
 \caption{Flow pattern of the four-Fermi coupling constant without
   $\bB$ (upper figure) and with $\bB$ (lower figure).  In the latter
   case chiral symmetry is always broken as a result of dimensional
   reduction.}
\label{fig:flow_mag}
\end{figure}

With this simple picture of chiral symmetry breaking we proceed to
finite temperature $T$ and quark chemical potential $\muq$.  More
specifically we search for the critical temperature (or chemical
potential) above which chiral symmetry is restored for a given
  $\lambda_\Lambda$.  The flow equation now receives medium
corrections, to wit,
\begin{align}\label{dtlambdafull}
 \partial_t \hlambda_k &= 2 \hlambda_k - \frac{\Nc\Nf\,\hlambda_k^2}
  {3\pi^2} \\
 &\quad \times  \biggl\{ 1 - (1-\partial_t) \bigl[n(k\!-\!\muq) +
  n(k\!+\!\muq) \bigr]\biggr\} \;, \notag
\end{align}
where $n(k)=1/(\rme^{\beta k}+1)$ is the Fermi-Dirac distribution
function.  Assume now that the initial coupling exceeds the fixed
point coupling at vanishing temperature, i.e.,
$\hlambda_\Lambda>\hlambda_\ast$ with $\hlambda_\ast$ in
Eq.~\eqref{eq:critical_rgflat} and hence
$\beta_{\hlambda}|_{k=\Lambda}<0$.  we again make contact to the
mean-field analysis within a solution of the differential
equation~\eqref{dtlambdafull} at $\muq=0$.  We can readily integrate
it to write down the running coupling constant,
\begin{equation}
 \lambda_k = \frac{\lambda_\Lambda}{\displaystyle 1 +
  \frac{\Nc\Nf\lambda_\Lambda}{3\pi^2}\! \int_\Lambda^k \!\!\rmd k'\,
  k'\biggl( 1-\frac{2}{\rme^{\beta k'}\!\!+\!1} \!-\!
  \frac{2\beta k'\rme^{\beta k'}}{(\rme^{\beta k'}\!\!+\!1)^2} \biggr)} \;.
\label{eq:running_T}
\end{equation}
Here, because we are interested in $\Tc$ inferred from the Landau pole
associated with the flow of $k\to0$, we make an approximation of
$k\sim 0$ and $\Lambda\sim\infty$ in the matter part and evaluate the
integration as
\begin{equation}
 \int_\infty^0 \rmd k'\, k'\biggl( - \frac{2}{\rme^{\beta k'}+1}
  - \frac{2\beta k'\rme^{\beta k'}}{(\rme^{\beta k'}+1)^2} \biggr)
 = \frac{\pi^2 T^2}{2} \;.
\end{equation}
Then, the running coupling~\eqref{eq:running_T} can be well
approximated as
\begin{equation}
 \lambda_k = \frac{\lambda_\Lambda}{\displaystyle
  1+\frac{\Nc\Nf\,\lambda_\Lambda}{6\pi^2}
  \bigl(k^2-\Lambda^2+\pi^2 T^2\bigr)} \;,
\end{equation}
from which the Landau pole position leads to the condition for the
critical temperature;
\begin{equation}
 \Tc = \sqrt{\frac{\Lambda^2}{\pi^2} - \frac{6}{\Nc\Nf\,\lambda_\Lambda}} \;.
\end{equation}
With the cutoff replaced as $\Lambda^2\to 3\Lambda^2$, this is
precisely consistent with the mean-field result in
Eq.~\eqref{eq:Tc-mf-zeroB}.

\subsection{Magnetic Catalysis}

We introduce finite $\bB$ using Eq.~\eqref{eq:Landau} and inserting
appropriate spin projection operators in the trace of the Dirac
indices.  In the LLLA we can easily understand the Magnetic Catalysis
from the flow pattern.  At $T=0$ the flow equation in the strong $\bB$
limit takes the following form;
\begin{equation}
 \partial_t \lambda_k = -\frac{\Nc}{2\pi^2}\sum_f
  |q_f B|\, \lambda_k^2
\label{eq:beta_B}
\end{equation}
with $t=\ln(k/\Lambda)$, which we can solve immediately as
\begin{equation}
 \lambda_k = \frac{\lambda_\Lambda}{\displaystyle
  1+\frac{\Nc\, \lambda_\Lambda}{2\pi^2}
  \sum_f|q_f B|\, t} \;.
\label{eq:running_B}
\end{equation}
Because $t=\ln(k/\Lambda)$ ranges from $0$ ($k=\Lambda$) to $-\infty$
($k=0$), the denominator of Eq.~\eqref{eq:running_B} inevitably hits
the Landau pole in the flow of $k$ in sharp contrast to the zero-$\bB$
case.  The existence of the Landau pole implies that the chiral
symmetric state is always unstable and so chiral symmetry should be
broken in the LLLA.\ \ This is nothing but the manifestation of the
Magnetic Catalysis.

In the same way as previously, we can understand this from the
behavior of the $\beta$-function of $\lambda_k$.  In contrast to
Eq.~\eqref{eq:beta} the mass dimension of $\lambda_k$ is now balanced
by not $k^2$ but $|q_f B|$ because of the dimensional reduction.
Therefore, the first term in Eq.~\eqref{eq:beta}, $2\hat{\lambda}_k$,
does not appear, and the behavior of the $\beta$-function changes as
sketched in Fig.~\ref{fig:flow_mag}.  More specifically, introducing a
dimensionless coupling by
$\lambda_k^{\rm (B)}\equiv(\sum_f|q_f B|/2\pi^2)\lambda_k$, the flow
equation reads,
\begin{equation}
 \beta_{\lambda^{\rm (B)}}\equiv-\Nc\bigl(\lambda_t^{\rm (B)}\bigr)^2 \;.
\label{eq:flow_dr}
\end{equation}
It is obvious from this that the flow goes to infinity regardless of
the initial point of $\lambda_\Lambda$.  From the
solution~\eqref{eq:running_B} we can easily locate the Landau pole at
$\ln(k_0/\Lambda)=-2\pi^2/(\Nc\lambda_\Lambda \sum_f|q_f B|)$, that
characterizes the typical scale for the condensate, i.e.\
\begin{equation}
 M_0 \propto k_0 = \Lambda\,\exp\biggl(-\frac{2\pi^2}
  {\Nc \lambda_\Lambda \sum_f|q_f B|}\biggr) \;,
\end{equation}
which gives an estimate for the chiral condensate up to an overall
coefficient.  We note that the exponential factor is identical with 
that in Eq.~\eqref{eq:mag_mass} since it does not depend on 
$\Lambda$ which carries the scheme-dependence. It does depend on 
$\sum_f|q_f B|\lambda_\Lambda$ which is already dimensionless.

\subsection{Finite temperature and density}

It is an interesting question how the chiral phase transition at
finite $T$ and/or $\muq$ is possible even in the LLLA in the RG
formulation.  In other words, the question is how the Magnetic
Catalysis is lost and chiral restoration becomes possible when a
finite $T$ and/or $\muq$ is turned on.  The running coupling constant
has extra contributions at finite $T$ as
\begin{equation}
 \begin{split}
 \lambda_k &= \lambda_\Lambda \Biggl[ 1 - \frac{\Nc\lambda_\Lambda}{2\pi^2}
  \sum_f |q_f B| \int_k^\Lambda \frac{\rmd k'}{k'} \\
 &\qquad\qquad\times \biggl( 1 - \frac{2}{\rme^{\beta k'}+1}
  - \frac{2\beta k'\,\rme^{\beta k'}}{(\rme^{\beta k'}+1)^2} \biggr)
  \Biggr]^{-1} \;.
 \end{split}
\label{eq:running}
\end{equation}
As we did before, we can approximate $k\simeq 0$ in the integration
range as long as we are interested in $\Tc$ only, and also take
$\Lambda\simeq\infty$, which makes the last term in the integration
part in Eq.~\eqref{eq:running} as
\begin{equation}
 \int_0^\infty \frac{\rmd k'}{k'}\,
  \frac{2\beta k'\,\rme^{\beta k'}}{(\rme^{\beta k'}+1)^2}
 = 1 \;.
\end{equation}
The first and the second terms are more interesting than the third
one.  The Magnetic Catalysis stems from the IR singularity from
$\int_k^\Lambda \rmd k'/k'=-\ln(k/\Lambda)$, while the finite-$T$
contribution exactly cancels the IR singularity as explicitly checked
as
\begin{equation}
 \int_0^\Lambda \frac{\rmd k'}{k'} \biggl(1\!-\!
  \frac{2}{\rme^{\beta k'}+1}\biggr)
 = \int_0^{\beta\Lambda} \frac{\rmd x}{x}\frac{\rme^x - 1}{\rme^x + 1}
  \simeq \ln( 1.13 \beta\Lambda ).
\end{equation}
This recovers the same expression for $\Tc$ as Eq.~\eqref{eq:Tc-mf-B}
up to the prefactor that again depends on the regularization scheme;
from the zero of the denominator in Eq.~\eqref{eq:running} we can
infer in the LLLA,
\begin{equation}
 \Tc = 0.42\Lambda \exp\biggl(-\frac{2\pi^2}{\Nc\lambda_\Lambda\,
  \sum_f |q_f B|} \biggr) \;.
\end{equation}
We can relax the LLLA and solve $\lambda_k$ for arbitrary $\bB$ at
finite $T$ and $\muq$ numerically.  We can write the running coupling
constant down as follows;
\begin{align}
 &\lambda_k = \lambda_\Lambda \Biggl[1-\frac{\Nc\lambda_\Lambda}{2\pi^2}
  \sum_f |q_f B| \int_k^\Lambda\frac{\rmd k'}{k'} \notag\\
 &\:\:\times \sum_{n=0}^{N_{k,B}}
  \alpha_n \sqrt{1-\frac{2|q_f B|n}{k^{\prime 2}}}
  \Bigl\{ 1 - n(k'\!-\!\muq) - n(k'\!+\!\muq) \notag\\
 &\qquad + k\partial_k \bigl[
  n(k'\!-\!\muq) + n(k'\!+\!\muq) \bigr] \Bigr\} \Biggr]^{-1} \;.
\end{align}
In this expression $n$ can take a positive integer
up to $N_{k,B}=\theta_g[k^2/(2|q_f B|)]$.  From the zero of the
denominator, the critical coupling constant $\lambda_\ast$ is fixed as
a function of $T$, $\muq$, and $B$ in unit of $\Lambda$, that is,
\begin{align}
 & \frac{1}{\lambda_\ast} = \frac{\Nc}{2\pi^2}\sum_f|q_f B|\int_0^1
  \frac{\rmd x}{x} \sum_{n=0}^{N_{x\Lambda,B}} \alpha_n
  \sqrt{1-\frac{2|q_f B|n}{x^2 \Lambda^2}} \notag\\
 &\quad \times \Biggl( 1 - \frac{1}{\rme^{\beta(x\Lambda-\muq)}\!+\!1}
   - \frac{1}{\rme^{\beta(x\Lambda+\muq)}\!+\!1} \notag\\
 &\quad\qquad  - \frac{\beta x\Lambda\,\rme^{\beta(x\Lambda-\muq)}}
  {(\rme^{\beta(x\Lambda-\muq)}\!+\!1)^2}
   - \frac{\beta x\Lambda\,\rme^{\beta(x\Lambda+\muq)}}
  {(\rme^{\beta(x\Lambda+\muq)}\!+\!1)^2} \Biggr) \;.
\label{eq:critical_full}
\end{align}

\begin{figure}
\includegraphics[width=0.5\textwidth]{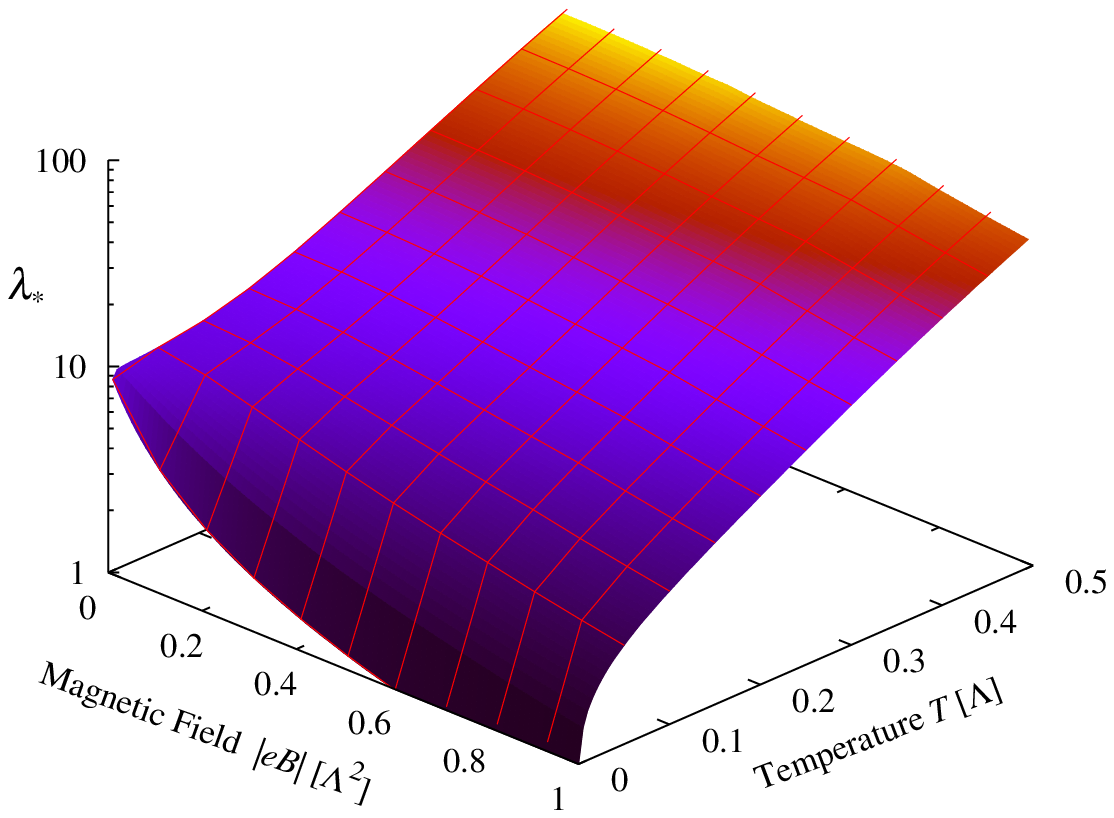}\\
\includegraphics[width=0.5\textwidth]{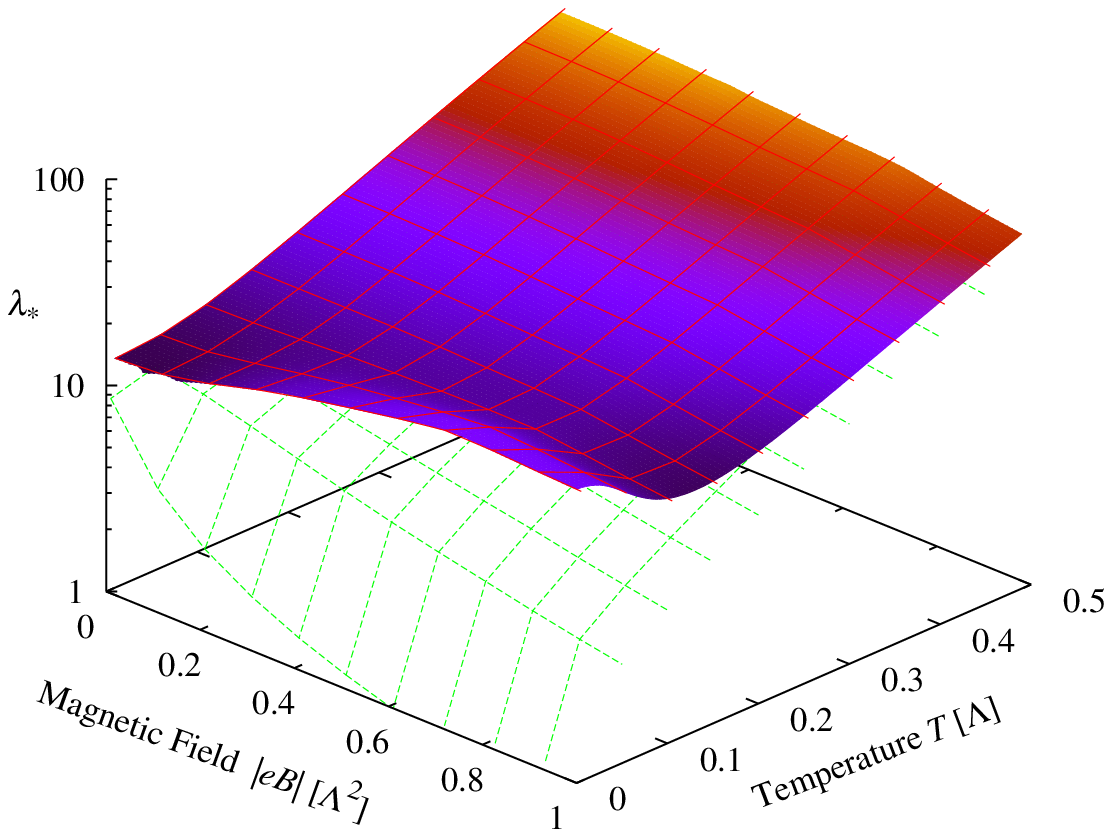}
\caption{Critical coupling constant $\lambda_\ast$ of chiral symmetry
  breaking and restoration as a function of the temperature $T$ and
  the magnetic field $|eB|$ in unit of $\Lambda$ for zero quark
  chemical potential $\muq=0$ (upper) and $\muq=0.3\Lambda$ (lower;
  with zero-density results shown for reference).}
\label{fig:lambda}
\end{figure}

The numerical results are shown in Fig.~\ref{fig:lambda} at zero
density (upper) and at finite density (lower).  We here like to point
out that the numerical results are entirely smooth.  This nice feature
is attributed to the choice of the optimized cutoff function.  In
fact, the Landau level summation stops at
$n=N_{k,B}$ but the contribution in the vicinity of this upper bound
is vanishingly small due to the weight
$\sqrt{1-2|q_f B|n/k^{\prime 2}}$.  In other words, the optimized
form~\eqref{eq:rk} is a small regularization function, though it
involves Heaviside's step function.

In the upper panel of Fig.~\ref{fig:lambda} we can confirm the
Magnetic Catalysis from the behavior of decreasing $\lambda_\ast$
toward zero with increasing $|eB|$ at $T=0$.  In view of the upper
panel of Fig.~\ref{fig:lambda}, however, the Magnetic Catalysis,
$\lambda_\ast=0$, is slowly reached for asymptotically large $\bB$
only.  We already mentioned that the LLLA is exact for
$2|qB|>\Lambda^2$.  In fact, the curve at $T\simeq0$ in
Fig.~\ref{fig:lambda} is not the result at strict zero-$T$ but at
$T=10^{-5}\Lambda$ for numerical stability.  As soon as finite $T$ is
introduced, $\lambda_\ast$ is significantly pushed up and the Magnetic
Catalysis can occur only in the limit of $\bB\to\infty$, though
$\lambda_\ast=0$ for any $\bB$ at strict zero-$T$.

It is also evident from the lower panel of Fig.~\ref{fig:lambda} that
the Magnetic Catalysis is lost immediately at finite $\muq$ and
$\lambda_\ast$ never approaches zero even at $T=0$ then.  More
interestingly, we can see in the figure that $\lambda_\ast$ increases,
meaning that chiral symmetry breaking weakens, with increasing $\bB$
contrary to the trend at $\muq=0$.  These are quite suggestive results
implying that the modification in the QCD phase boundaries induced by
$\bB$ weakens at higher temperature, and $\bB$ would rather favor
chiral restoration for cold and dense quark matter.

\section{Non-local Vertex and Resummation}
\label{sec:meson}

So far, we have worked in the leading order of the $1/\Nc$-expansion
which amounts to neglecting the diagrams (b),(c),(d) in
Fig.~\ref{fig:flow}.  Once they are included in the calculation, the
overall coefficient is modified from $2\Nc$ to $2\Nc+\frac{3}{4}$,
leaving aside the complexity due to isospin symmetry breaking by
$\bB$. 

Hence, in practice, the sub-leading contributions have the minor effect
of changing the overall factor.  Strictly speaking, however, this is
not the end of the story.  The running coupling constant has an
interpretation in terms of the meson propagation in the intermediate
state as illustrated in the schematic
representation~\eqref{eq:schematic}.  Because we defined $\lambda_k$
from the four-point vertex function with zero momentum insertion from
external legs, we could describe the meson propagation only for zero
momentum, as explicitly indicated in Eq.~\eqref{eq:dt1ovlambda}.  Such
a treatment is correct for (a), but not adequate for (b), (c), and
(d).  In this section we will discuss qualitative effects of the
momentum-dependent (i.e.\ non-local) vertices mediated by the meson
propagation.

\subsection{Partial resummation}

Using the graphical representation~\eqref{eq:schematic} we can express
the diagram (d) as a box-type diagram as follows;
\begin{equation}
 \includegraphics[width=0.4\columnwidth]{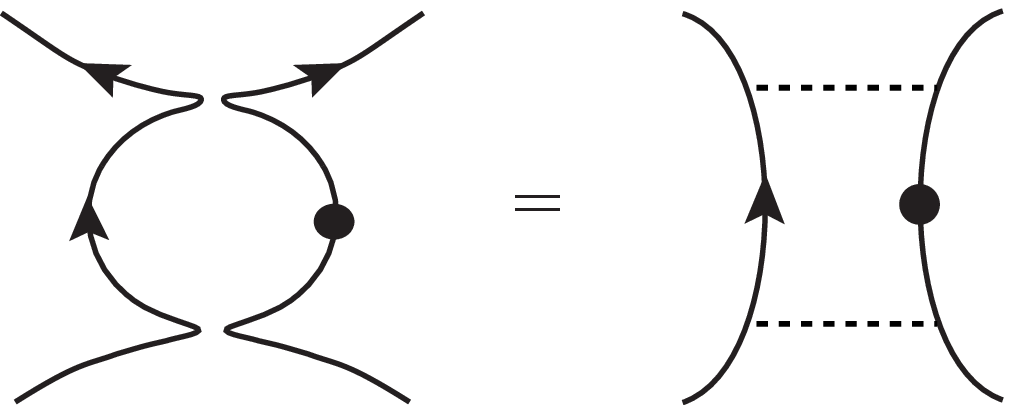}
\end{equation}
Here, it is clear that two meson propagators must carry the internal
loop momentum and $\lambda_k$ in the diagram should be replaced by the
momentum-dependent coupling $\lambda_k(p)$. For the diagrams (b) and
(c) as well as (a) we can utilize a similar representation with
non-local vertices.

One might have thought of a different type of the box diagram, that
is, the one with the $s$- and $t$-channels flipped.  However, such a
diagram does not give an additional contribution because, as we have
explained below Eq.~\eqref{eq:deriv}, the color structure is
decomposed into two pieces.  We are now handling the part of
$\delta^{\alpha\beta}\delta^{\gamma\delta}$ and the flipped diagram
belongs to another part of
$\delta^{\alpha\delta}\delta^{\beta\gamma}$.  This is a crucial
difference from the box diagram in terms of flavor indices
encountered, for example, in the QED analysis~\cite{Aoki:1996fh}.

To take account of these effects properly, therefore, we should
use the momentum-dependent coupling constant and it is natural to
anticipate the following replacement;
\begin{equation}
 \begin{split}
  &\lambda_k =
  \frac{\lambda_\Lambda}{1-\lambda_\Lambda[\Pi_k(0)-\Pi_\Lambda(0)]} \\
  &\Rightarrow\quad \lambda_k(p) =
 \frac{\lambda_\Lambda}{1-\lambda_\Lambda[\Pi_k(p)-\Pi_\Lambda(p)]}  \;. 
 \end{split}
\label{eq:mom_dep}
\end{equation}
This is nothing but the coupling arising from a full
momentum-dependent bubble resummation and can be achieved within
2PI-approximations to the flow~\cite{Blaizot:2010zx}. 

Within such a momentum-dependent approximation we should discuss anew
the validity of the LLLA.\ \ We can imagine the complexity from a
simple fact that the energy dispersion of charged quarks is Landau
quantized, while neutral mesons consisting of those quarks do not feel
$\bB$ at all and thus no Landau quantization occurs if they are
tightly bound.  We can indeed understand this from
Eq.~\eqref{eq:mom_dep}; the point is that $\Pi_k(p)$ in
Eq.~\eqref{eq:mom_dep} is the polarization in the presence of the IR
cutoff $k$, that is, the IR modes with $p^2+2|q_f B|n < k^2$ are
massive and do not run in the loop, which makes $\Pi_k(p)$ dependent
on $k$.  As long as $|q_f B|$ is well smaller than $\Lambda^2$, the
magnetic field cannot resolve the internal structure of bound-state
system and the LLLA is not sufficient.  In this case the apparent
suppression of quark propagator by $|q_f B|$ is canceled by the energy
denominator in the polarization calculation.  On the other hand, if
$|q_f B|$ is greater than $\Lambda^2$, all non-zero Landau modes
($n\neq0$) are outside of the IR cutoff region and only the zero-mode
can depend on $k$.  Therefore, only the zero-mode survives in
$\Pi_k(p)-\Pi_\Lambda(p)$.  Roughly speaking, the transverse motion of
neutral mesons gets frozen if the magnetic field strength is larger
than the inverse squared of the meson size and this could be regarded
as $B$-induced dissociation or deconfinement.  Such a transitional
change of meson structure as a function of $B$ is a very interesting
theoretical question beyond the scope of this present work.

Let us just assume that $|q_f B|$ is sufficiently large and the LLLA
holds in the computation of $\Pi_k(p)-\Pi_\Lambda(p)$.  Then, after
some calculations, we find,
\begin{equation}
 \begin{split}
 &\hspace{-0.7em} \Pi_k(p)-\Pi_\Lambda(p) \\
 \simeq &-\frac{\Nc}{2\pi^2}\sum_f |q_f B|
  \biggl[ \ln(k/\Lambda) + \frac{|p_z|}{2k} + \frac{3p_4^2}{8k^2}
  + \frac{p_z^2}{8k^2} \biggr] \;,
 \end{split}
\label{eq:pol}
\end{equation}
where we expanded the results in terms of $p_4/k$ and $p_z/k$ and took
$\Lambda \gg p_4,\,p_z$ to neglect $p_4/\Lambda$ and $p_z/\Lambda$.
We can immediately confirm that we can recover the running
coupling~\eqref{eq:running_B} by plugging Eq.~\eqref{eq:pol} into
Eq.~\eqref{eq:mom_dep} at zero momenta.  Because $p_z$ is the loop
momentum in the diagrams (b)--(d) which is cutoff at $k$ by the quark
propagator, $p_z$ and $p_4$ are certainly smaller than $k$.  Since the
introduction of $k$ explicitly breaks Lorentz symmetry,
Eq.~\eqref{eq:pol} is a function of not only $p_4^2+p_z^2$ but also
$|p_z|$ independently.  We have introduced $r_k(\bp)$ in the quark
propagator to suppress the IR modes with $|\bp|<k$ and, somehow, the
meson propagator should have the IR cutoff generated dynamically.
Then, the spatial momentum of meson is shifted up by the IR cutoff
$k$.  To take account of this cutoff effect qualitatively, it would
suffice to keep the linear term only in Eq.~\eqref{eq:pol}.  The
running coupling with momentum dependence is then expressed as
\begin{equation}
 \lambda_k(p) = \frac{\lambda_k}{\displaystyle 1+
  \frac{\Nc\lambda_k}{4\pi^2}\sum_f |q_f B|\,\frac{|p_z|}{k}} \;.
\label{eq:mom_dep2}
\end{equation}
We can then perform the momentum integration.  For qualitative
discussions we can treat $\lambda_k(\bar{p})$ with some typical
$\bar{p}$ in Eq.~\eqref{eq:mom_dep2}.

In the vicinity of chiral phase transition, $\lambda_k$ in
$\lambda_k(p)$ becomes larger and larger as $k$ decreases, which is a
flow pattern implied from Eq.~\eqref{eq:beta_B}.  Na\"{i}vely, one
might have thought that the effect of the diagrams (b)--(d) is just to
replace the overall factor in the $\beta$-function, but this is not
the case once the momentum dependence in the running coupling constant
is taken into account.  As is clear from Eq.~\eqref{eq:mom_dep2}, we
can see,
\begin{equation}
 \lambda_k(\bar{p}) \to \text{(const.)} \qquad
 (\lambda_k \to \infty;\; k \to 0) \;.
\end{equation}
This means that the contribution from the diagram (a) which is
proportional to $\lambda_k^2$ overwhelms (b)--(d) which are
proportional to $\lambda_k(\bar{p})\lambda_k$ and
$\lambda_k(\bar{p})^2$, respectively, near the chiral phase
transition.  Therefore, our analysis of (a) in the previous section is
not changed by the inclusion of (b)--(d) qualitatively.

To go into more quantitative studies, however, it is inadequate to use
$\lambda_k$ that is solved previously, but we need to solve the flow
equation self-consistently including the non-local vertices.

\subsection{Self-consistent resummation} 

Finally we advance our truncation scheme for taking account the
non-local vertices $\lambda_k(p)$.  The result for $\lambda_k(p)$
discussed previously stems from a standard resummation of bubble
diagram in the $s$-channel or the leading-order contribution in the
$1/\Nc$ expansion.  Now we utilize the flow for going beyond this
approximation.  For that purpose we parameterize the full $s$-channel
coupling $\lambda_k(p)$ as
\begin{equation}
 \lambda_k(p) = \frac{\bar{\lambda}_k(p)}
  {1-\bar{\lambda}_k(p)\Delta\Pi_k(p)}
\label{eq:fulls}
\end{equation} 
with an unknown function $\bar{\lambda}_k(p)$.  One can understand
this Ansatz as a running coupling~\eqref{eq:mom_dep} with shifted
renormalization.  That is, one can immediately reach
Eq.~\eqref{eq:fulls} by the following replacement in
Eq.~\eqref{eq:mom_dep};
\begin{equation}
 \frac{1}{\lambda_k} + \Delta\Pi_\Lambda(p)
  \to \frac{1}{\bar{\lambda}_k(p)} \;.
\end{equation}
We note that we have to promote the right-hand side as a
momentum-dependent quantity to take account of the full resummation
including (b)--(d) in addition to (a).  We can fix the normalization
at vanishing momentum as
\begin{equation}
 \bar{\lambda}_k(0) = \lambda_k(0) = \lambda_k \;.
\label{eq:lambdak0}
\end{equation}
The flow of $\lambda_k(p)$ is, in principle, given by the flow
equation with diagrams having momentum insertion from the external
legs.  For the derivation of the flow of $\bar{\lambda}_k(p)$ we can
take the derivative on Eq.~\eqref{eq:fulls} to find,
\begin{equation}
 \frac{\partial_k \bar{\lambda}_k(p)}{\bar\lambda_k(p)}
 = \frac{\partial_k \lambda_k(p)}{\lambda_k(p)}
  \frac{1}{1+\lambda_k(p)\Delta\Pi_k(p)}
  -\lambda_k(p)\partial_k \Delta\Pi_k(p) \;.
\label{eq:fullbars}
\end{equation}
For the solution of the flow $\partial_k \bar{\lambda}_k(p)$ in
Eq.~\eqref{eq:fullbars} we have to insert the flow of
$\partial_k \lambda_k(p)$ in a given approximation.  A fully
self-consistent flow results in loops with momentum-dependent
four-point vertex which would require numerical computations of the
Dyson-Schwinger type.

In the present work we invoke further approximations motivated by the 
physics under discussion.  This also facilitates the task of solving
the flows and to uncover the underlying mechanisms.

The coupling parameter $\bar{\lambda}_k(p)$ encodes the pion effects
beyond the leading $s$-channel resummation.  These effects, as is
already the case in the previous subsection for the $s$-channel
resummation, are suppressed for non-zero internal momenta.  This
structure enables us to solve the flow for $\bar{\lambda}_k(p)$ within
a derivative expansion in the $s$-channel momentum $p$.  It also helps
that the successive integration in terms of $k$ already includes the
momentum-dependence of $\bar{\lambda}_k(p)$ within the identification,
\begin{equation}
 \lambda_{k=0}(q) \simeq \lambda_{k=q}(0) \;.
\label{eq:ktop}
\end{equation}
Equation~\eqref{eq:ktop} also provides a self-consistency check of the
current derivative expansion.  In the present work we take the
lowest-order derivative expansion as
$\bar{\lambda}_k(p)\simeq \bar{\lambda}_k(0)=\lambda_k$ [see
Eq.~\eqref{eq:lambdak0}] and, analogously to the flow
equation~\eqref{eq:dtlambda}, we arrive at
\begin{equation}
 \partial_k \lambda_k = \sum_{i=a,b,c,d}{\rm Diag}^{(i)} \;,
\label{eq:newdotlambda}
\end{equation}
where ${\rm Diag}^{(i)} $ with $i=a,b,c,d$ stands for the momentum
integrations corresponding to the diagrams (a)--(d).  With the
coupling $\lambda_k(p)$ put down in Eq.~\eqref{eq:fulls} the diagrams
take the form;
\begin{align}
 &{\rm Diag}^{(a)} = -8\Nc \sum_{f=u,d} \int
  \frac{\rmd^4 p}{(2\pi)^4}\; \lambda_k^2 \; I_k(p_4,\bp) \;,\\
 &{\rm Diag}^{(b)}+{\rm Diag}^{(c)} \notag\\
 &\quad = -4\sum_{f=u,d}\int
  \frac{\rmd^4 p}{(2\pi)^4} \;\frac{\lambda_k^2}
  {1-\lambda_k\Delta\Pi_k(p)} \; I_k(p_4,\bp) \;,\\
 &{\rm Diag}^{(d)} = \sum_{f=u,d}\int\frac{\rmd^4 p}{(2\pi)^4}\;
  \frac{\lambda_k^2}{[1-\lambda_k\Delta\Pi_k(p)]^2} \;
  I_k(p_4,\bp) \;.
\label{eq:loopi}
\end{align}
with the common integrand given by
\begin{equation}
 I_k(p_4,\bp) = \frac{\partial_k r_k(\bp)\,
  \bp^2 [1+r_k(\bp)]}{\{p_4^2+\bp^2[1+r_k(\bp)]^2\}^2} \;.
\end{equation}
For the diagram (a) the $s$-channel momentum is the external one for
both couplings and hence $\lambda_k^2$ appears in the integration just
in the same way as in Eq.~\eqref{eq:dtlambda}.  For the diagrams (b)
and (c) one of two couplings has zero momentum in the $s$-channel and
the other has the loop momentum $p$.  Finally, for the diagram (d), we
have $p$ for both vertices.

It should be noted that we have neglected isospin symmetry breaking in
the above expressions.  In other words, we here introduced
simplification by setting the electric charges of $u$-quarks and
$d$-quarks to be equal; $q_u = q_d$.  Without this, the polarization
$\Pi_k(p)$ is dependent on which flavors of quarks each diagram
involves and thus $\lambda_k(p)$ should be also flavor dependent.  As
a result, for example, the charged pion propagations are suppressed
and the contributions from (b) and (c) are vanishing due to
cancellation between the $\pi^0$ and $\sigma$ processes.  We could
have discussed all these complexities here, but they are not the main
subject of the present work.  We are only interested in formulating
the resummation procedure here and looking over qualitative effects on
the flow.  Though we have to leave from real electromagnetism for the
moment, we shall stick to simplicity with $q_u=q_d$.

Now we have a closed flow equation for $\lambda_k$.  It takes into
account the scale dependence of the coupling at vanishing momentum and
goes beyond the standard bubble resummation.  The current
approximation fully restores RG-invariance; the coupling
$\lambda_k(p)$ is invariant under a RG rescaling and looses all
reference to the coupling at the cutoff, i.e.,
$\partial_{\lambda_\Lambda}\lambda_{k=0}=0$.

Let us now discuss how the $\beta$-function is modified by this
prescription of resummation.  For this purpose we will again focus on
the calculation in the strong $\bB$ limit which enables us to use the
LLLA for analytical calculations.  In the LLLA, as we have seen,
$\Delta\Pi_k(p)$ can be approximated as
$\Delta\Pi_k(p)\simeq-(\Nc/4\pi^2)\sum_f|q_f B| (|p_z|/k)$.  Since the
momentum dependence is such simple, one can easily integrate each
${\rm Diag}^{(i)}$ with non-local vertices.

\begin{figure}
\includegraphics[width=\columnwidth]{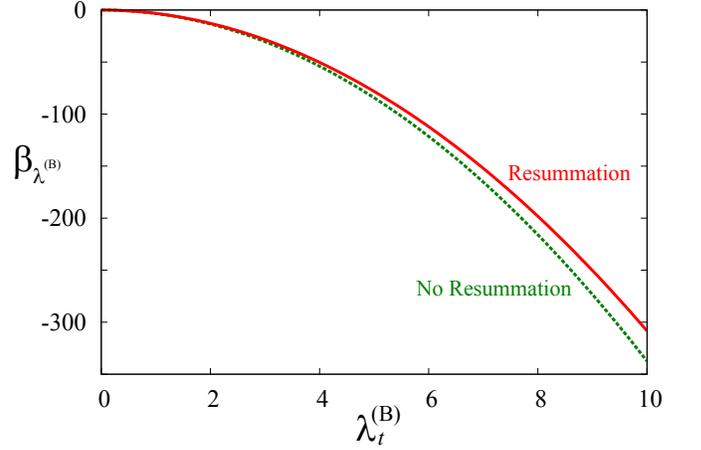}
\caption{Flow patterns of the four-Fermi coupling constant without
  resummation shown by the dotted line (in which the overall
  coefficient is $2\Nc+\frac{3}{4}$) and with resummation shown by
  the solid line.}
\label{fig:flow_mag3}
\end{figure}

After the $p_z$-integration, with this resummation procedure, the
$\beta$-function changes from Eq.~\eqref{eq:flow_dr} into
\begin{equation}
 \begin{split}
 &\partial_k\lambda_t^{\rm (B)}=-\Nc \bigl(\lambda_k^{\rm (B)}\bigr)^2 \\
 &\qquad\qquad -\frac{\lambda_k^{\rm (B)}}{\Nc}
  \ln\Bigl(1+\tfrac{3}{2}\lambda_k^{\rm (B)}\Bigr)
  +\frac{(\lambda_k^{\rm (B)})^2}
  {8\bigl(1+\tfrac{3}{2}\lambda_k^{\rm (B)}\bigr)} \;.
 \end{split}
\end{equation}
The behavior of the right-hand side is shown in
Fig.~\ref{fig:flow_mag3}.  We can see that the resummation has a minor
effect as stated in qualitative discussions in the previous
subsection.  The resummation tends to disfavor chiral symmetry
breaking slightly, which is consistent with the intuition that meson
fluctuations would rather restore chiral symmetry.

\section{Conclusion}
\label{sec:conclusion}

We have investigated the RG flow pattern of the four-Fermi coupling
constant $\lambda_k$ and clarified the intuitive picture of the chiral
phase transition at zero and finite temperature and baryon density.
We have applied the formalism to study the effects of the homogeneous
magnetic field.  In the RG analysis it is extraordinarily simple to
understand the phenomenon called the Magnetic Catalysis.  If the
magnetic field is strong enough, the transverse dynamics of charged
particles is completely frozen and the dimensional reduction
effectively occurs.  Then, with the IR cutoff scale $k$, the
transverse phase-space factor $k^2$ is replaced by the Landau
degeneracy $|q_f B|/(2\pi)$, which changes the flow pattern
significantly.

For dimensional reasons, in the absence of an external magnetic field
$\bB$, a dimensionless coupling constant is given by $\lambda_k k^2$,
which tends to go to zero as $k$ decreases unless the initial
$\lambda_\Lambda$ exceeds a critical value.  In contrast, if strong
magnetic fields $\bB$ are applied, the dimensionless combination of
coupling constant is rather characterized by
$(\sum_f|q_f B|/2\pi^2)\lambda_k$.  For this combination one can show
that its RG flow leads to a diverging coupling however small the
initial $\lambda_\Lambda$ is.  This is the simple RG-picture of the
Magnetic Catalysis.

Our quantitative studies in numerical calculations have made it clear
that the dimensional reduction, however, is a sensible approximation
only when $B$ is unrealistically large and the temperature $T$ and the
chemical potential $\muq$ are sufficiently close to zero.  The
logarithmic singularity that is responsible for the Magnetic Catalysis
is exactly canceled by another one from the matter parts at finite
$T$ and/or $\muq$.

Finally, we have included the effects of non-local vertices mediated
by the meson propagation with finite momentum.  The resummation
procedure argued in this work thus deals with separate contributions
from meson loop effects and contains non-trivial contributions through
the RG flow; see Ref.~\cite{Blaizot:2010zx} for a similar kind of
resummation.  The related set of diagram has no qualitative impact on
the location of the chiral phase transition.  The present novel
approximation to the NJL-model takes into account the back-reaction of
intermediate mesonic states and hence relates to the full quark-meson
flow in (P)QM models; see e.g.\ 
\cite{Skokov:2010wb,Herbst:2010rf,Skokov:2011ib,Herbst:2012ht}.  We
hence expect the present approximation to be sensitive to the critical
properties of the phase transition.  The discussion of critical
properties is deferred to future extensions of the present work.

Although it was not quite necessary in the present work, it would be
an intriguing question how to describe the neutral meson as charged
quark composite in external $\bB$.  Our RG analysis clearly shows that
the lowest Landau level approximation works for the neutral meson
propagation only when $B$ is greater than the cutoff scale $\Lambda$
that roughly corresponds to the transverse meson size inverse.  This
means that the quark dissociation or deconfinement could be induced by
the $\bB$ effects, which would offer us a new opportunity to shed
light on physics of quark confinement.  Besides, it is an important
but poorly understood problem to clarify the physical meaning of the
conventional criterion for quark confinement using the Polyakov loop
at strong $\bB$, which would help us with understanding
confinement/deconfinement at high density and hopefully more detailed
characteristics of Quarkyonic Matter~\cite{McLerran:2007qj}.  These
are all interesting future extensions of the present work.

\acknowledgments
The authors thank Jens~Braun, Holger~Gies, Kouji~Kashiwa,
Anton~Rebhan, Andreas~Schmitt for discussions.


\bibliographystyle{apsrev4-1}
\bibliography{magnetic}

\end{document}